\documentclass[twocolumn,prb,preprintnumbers]{revtex4}
\usepackage{amssymb,graphicx}

\begin{document}

\title{First-principles
study on physical properties of a single ZnO monolayer with
graphene-like structure}

\author{Z. C. Tu}\email{tuzc@bnu.edu.cn}
\affiliation{Department of Physics, Beijing Normal University,
Beijing 100875, China}

\begin{abstract}
The elastic, piezoelectric, electronic, and optical properties of a
single ZnO monolayer (SZOML) with graphene-like structure are
investigated from the first-principles calculations. The phonon
dispersion curves contain three acoustic and three optical branches.
At $\Gamma$ point, the out-of-plane acoustic mode has an asymptotic
behavior $\omega (q)=Bq^2$ with $B=1.385 \times 10^{-7}$ m$^2$/s,
while two in-plane acoustic modes have sound velocities $2.801$ km/s
and $8.095$ km/s; the other three optical modes have frequencies 250
cm$^{-1}$, 566 cm$^{-1}$, and 631 cm$^{-1}$. The elastic and
piezoelectric constants are obtained from the relaxed ion model. It
is found that the SZOML is much softer than graphene, while it is a
piezoelectric material. The electronic band gap is 3.576 eV, which
implies that the SZOML is a wide band gap semiconductor. Many peaks
exist in the linear optical spectra, where the first peak at 3.58 eV
corresponds to the band gap of SZOML.
\preprint{J. Comput. Theor. Nanosci. 7, 1182-1186 (2010)}
\end{abstract}
\maketitle
\section{Introduction}
Many ZnO nanostructures such as nanowires, nanobelts, nanorings and
nanohelixes \cite{Huangmhsci,Panzw,Kongxysci} have been synthesized
in recent years, which are expected to be used to make novel
electronic devices such as field-effect transistors, nanogenerators
and so on. \cite{wangzljpc,Wangsci06} Viewed from the small scales,
these nanostructures are locally wurtzite type. In 2005, Claeyssens
\emph{et al.} \cite{Claeyssens05} found that the graphite-like
structure was more energetically favorable than the wurtzite
structure for a very thin ZnO film through the density functional
theory (DFT) calculations. They obtained the critical number of ZnO
monolayers in the thin film to be 18, beyond which the graphite-like
structure would transit to the bulk wurtzite structure
\cite{Claeyssens05}. Subsequently, the present author and Hu
\cite{TuHu06} pointed out that DFT calculations might overestimate
the interlayer interactions because the widely used DFT packages
cannot accurately describe the non-covalent bonds. \cite{Kohn98} The
critical number was estimated to be 3 based on some arguments for
the interlayer interactions between two ZnO monolayers and the DFT
calculations for a single ZnO monolayer (SZOML). \cite{TuHu06} In
2007, Tusche \textit{et al.}\cite{Tusche07} observed graphite-like
ZnO monolayers in the experiment and found that critical number to
be 3 or 4.

We expect that this new material might have good physical
properties. However, there is very few investigation on this topic
except our previous work \cite{TuHu06} on elastic constants and the
work \cite{BotelloCPL07,Botellonl08} on metallic edges and magnetic
behavior in ZnO nanoribbions. Here, we will address the elastic,
piezoelectric, electronic, and optical properties of an SZOML with
graphene-like structure by first-principles calculations (the ABINIT
package\cite{Gonzexcms}). The rest of this paper is organized as
follows: In section \ref{sec-stru}, we show an optimized structure
of the SZOML. In section \ref{sec-phon}, we calculate the phonon
dispersion relation. In section \ref{sec-piez}, we obtain the
elastic and piezoelectric constants based on the relaxed ion model.
In section \ref{sec-elec}, the electronic band structure and the
band gap are obtained through the DFT calculation and GW
approximation. In section \ref{sec-opt}, we show the results of the
optical dielectric functions. We summarize the main results and
infer the possible physical properties of single-walled ZnO
nanotubes based on these main results in section \ref{sec-condis}.

\section{Structure \label{sec-stru}}
\begin{figure}[!htp]
\includegraphics[width=7cm]{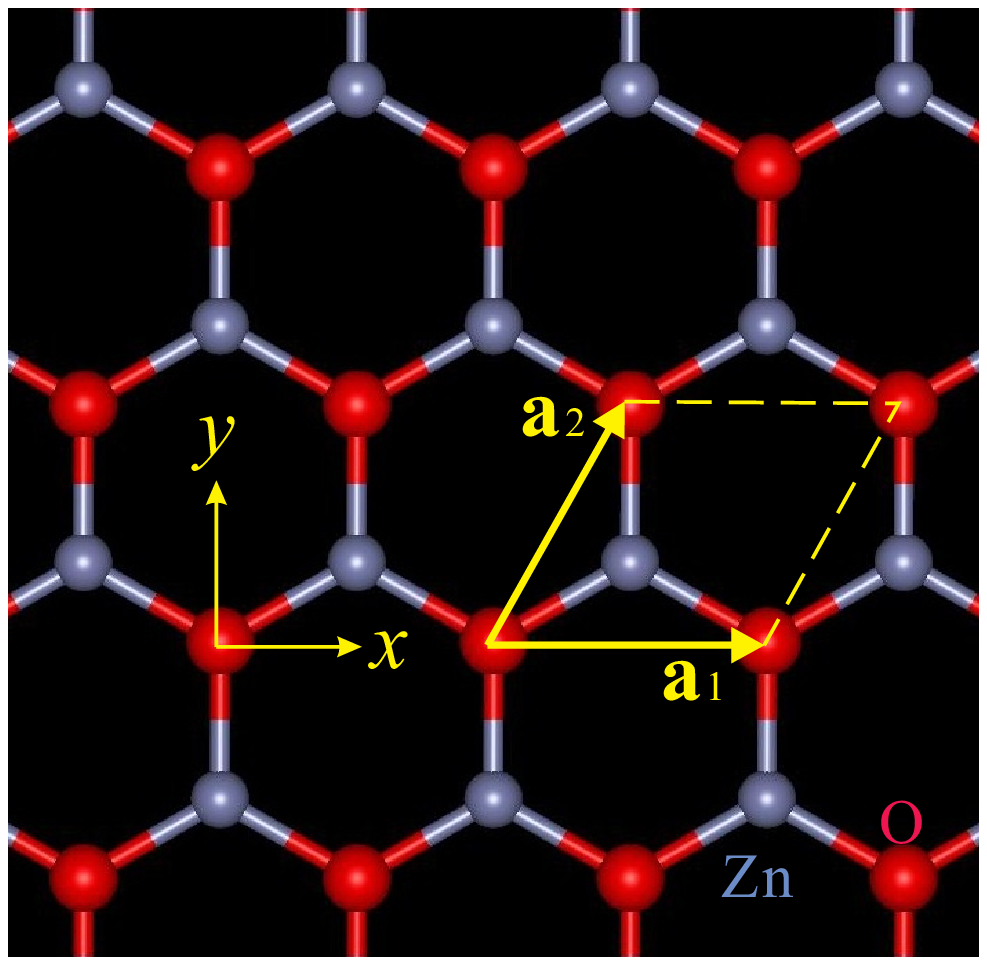}
\caption{\label{fmonolayer}(Color online) Graphene-like ZnO
monolayer. $\mathbf{a}_1$ and $\mathbf{a}_2$ are the two lattice
vectors.}\end{figure}

The DFT calculations \cite{Claeyssens05,TuHu06,Ligao07,Fisker09} and
the experiment \cite{Tusche07} reveal that an SZOML with
graphene-like structure is chemically stable. Its structure is
schematically depicted in Fig.~\ref{fmonolayer} where $\mathbf{a}_1$
and $\mathbf{a}_2$ are the two lattice vectors. Each unit cell
contains one Zn atom and one O atom. We optimize the structure by
taking Troullier-Martins pseudopotentials, \cite{Troullier}
plane-wave energy cutoff 50 Hartree, and $8\times8\times 1$
Monkhorst-Pack k-points \cite{MonkhorstPack} in Brillouin-zone. The
exchange-correlation energy are treated within the local-density
approximation in the Ceperley-Alder form \cite{CeperleyAlder} with
the Perdew-Wang parametrization. \cite{PerdewWang} We keep the
lattice constant in $z$-direction 40 bohr such that our result is
available for the SZOML. We obtain the two lattice constants in
$xy$-plane $|\mathbf{a}_1|=|\mathbf{a}_2|=6.062$ bohr (=3.208 \AA),
and correspondingly, the bond length 1.853 \AA, which is close to
the previous value \cite{TuHu06} obtained by the present author and
Hu with different ecut and k-points.

\section{Phonon dispersion relation \label{sec-phon}}
\begin{figure}[!htp]
\includegraphics[width=7cm]{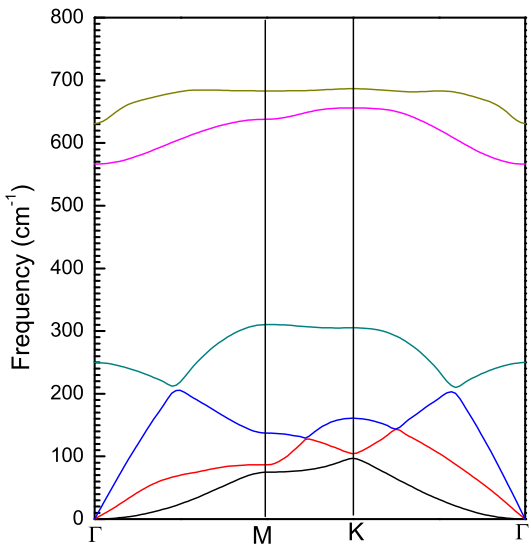}
\caption{\label{fphonon}(Color online) Phonon dispersion curves
plotted along high symmetry directions in the Brillouin
zone.}\end{figure}

The phonon dispersion relation $\omega =\omega(q)$ entirely reflects
the lattice dynamic behavior of SZOML, where $\omega$ and $q$ are
the phonon frequency and wave number, respectively. Here, we obtain
the phonon dispersion relation through the first-principles
calculations following the work \cite{GonzePRB97,Giannozzirmp01} by
Gonze and others and show the results in Fig.~\ref{fphonon}.

Three phonon dispersion branches originating from $\Gamma$ point of
the Brillouin zone correspond to the acoustic modes: an out-of-plane
mode (the lower branch), an in-plane tangential mode (the middle
branch) and an in-plane radial mode (the upper branch). Among them,
the phonon dispersion curve of the out-of-plane mode near $\Gamma$
point has an asymptotic behavior $\omega (q)=Bq^2$ with $B=1.385
\times 10^{-7}$ m$^2$/s. The in-plane tangential and radial modes
have linear behaviors near $\Gamma$ point. The sound velocities are
$v_t =2.801$ km/s and $v_r=8.095$ km/s, respectively, which are much
smaller than the corresponding velocities $15$ km/s and $24$ km/s
for graphene. \cite{Dresselhaus20} Here we should emphasize that one
cannot directly derive the in-plane rigidity and shear modulus from
the sound velocities $v_t$ and $v_r$ because the elasticity and
piezoelectricity of SZOML couples with each other for in-plane
deformations. \cite{BornHuang} But the piezoelectricity and the
out-of-plane mode are decoupled. Thus we can derive the bending
rigidity $k_c$ of SZOML from $B$. The relation between them is
\begin{equation}k_c=\sigma B^2,\end{equation}
where $\sigma=1.516\times 10^{-6}$ kg/m$^2$ is the density (i.e.,
mass per area) of SZOML. Then we obtain $k_c =0.182$ eV, which is
much smaller than the bending rigidity (1.62 eV)\cite{Tuoy08} of
graphene.

The remaining three branches are optical modes: one out-of-plane
mode and two in-plane modes. At $\Gamma$ point, their frequencies
are non-vanishing, and the corresponding values are 250 cm$^{-1}$,
566 cm$^{-1}$, 631 cm$^{-1}$, respectively.

\section{Elasticity and piezoelectricity \label{sec-piez}}
We have mentioned that the elastic constants of SZOML cannot be
directly derived from the phonon dispersion relation because the
coupling between elasticity and piezoelectricity. However, we can
directly calculate the elastic and piezoelectric constants by using
the ABINIT package. Because the SZOML has threefold rotation
symmetry around $z$-axis and a reflection symmetry with respect to
the $y$-axis as shown in Fig.~\ref{fmonolayer}, the elasticity and
piezoelectricity for zero external electric field can be expressed
in the matrix form \cite{Nyebook} as
\begin{equation}
\left[\begin{array}{c} \sigma_1\\ \sigma_2\\
\sigma_6\end{array}\right]=\left[\begin{array}{ccc} c_{11} & c_{12} &0\\
c_{12}&c_{11}&0\\
0&0&\frac{c_{11}-c_{12}}{2}\end{array}\right]\left[\begin{array}{c} s_1\\ s_2\\
s_6\end{array}\right].
\end{equation}
and \begin{equation} \left[\begin{array}{c} P_1\\
P_2-P_2^0\end{array}\right]=\left[\begin{array}{ccc}
0&0&-d_1\\ -d_1 & d_1 &0\end{array}\right]\left[\begin{array}{c} s_1\\ s_2\\
s_6\end{array}\right],
\end{equation}
respectively. In the above expressions, $\sigma_1$ and $\sigma_2$
are the normal stresses along $x$ and $y$ directions, respectively.
$s_1$ and $s_2$ are the normal strains along $x$ and $y$ directions,
respectively. $\sigma_6$ and $s_6$ are the in-plane shear stress and
shear strain, respectively. $P_2^0$ is the spontaneous polarization
in the $y$-direction. $P_1$ and $P_2$ are polarizations along $x$
and $y$ directions, respectively. $c_{11}$ and $c_{12}$ are two
independent elastic constants, while $d_1$ is the independent
piezoelectric constant. Because the thickness of SZOML is unknown,
we let it be implicit in $c_{11}$, $c_{12}$, and $d_1$. Based on the
relaxed ion model, \cite{BornHuang} we obtain $c_{11}=56.8$ eV/ZnO,
$c_{12}=40.3$ eV/ZnO, and $d_1=3.0$ $p_e$/ZnO from DFT calculations,
where $p_e=8.5\times 10^{-30}$ C m is the atomic unit of electric
dipole moment.

Using the elastic constants, we can derive the in-plane Young's
modulus ($Y$), shear modulus ($G$), and Poisson ratio ($\nu$) of
SZOML as follows:
\begin{eqnarray}
Y&=&c_{11}(1-c_{12}^2/c_{11}^2)=28.1~\mathrm{eV/ZnO},\label{eqyoung}\\
G&=&(c_{11}-c_{12})/2=8.2~\mathrm{eV/ZnO},\\
\nu&=&c_{12}/c_{11}=0.71.\label{eqpoisson}
\end{eqnarray}
The Young's modulus and shear modulus of SZOML are much smaller than
those of graphene, 115.4 and 49.6 eV/CC, respectively, while the
Poisson ratio of SZOML is much larger than that of graphene, 0.16.\cite{Tuoy08,notepration}

\section{Electronic band structure \label{sec-elec}}
\begin{figure}[!htp]
\includegraphics[width=7.5cm]{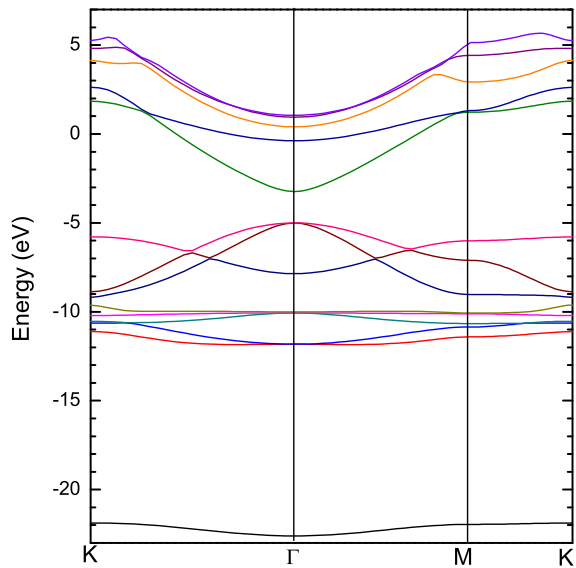}
\caption{\label{felecband}(Color online) Energy dispersion curves
for 9 valence and 5 conduction bands plotted along high symmetry
directions in the Brillouin zone.}\end{figure}

As a new material, we expect that SZOML has a good electronic
property. Its band structure is calculated within the DFT framework.
In Fig.~\ref{felecband}, we plot energy dispersion curves for 9
valence and 5 conduction bands along high symmetry directions in the
Brillouin zone. There are 1 low energy valence band, 5 middle energy
valence bands, and 3 high energy bands among the 9 valence bands.
The direct gap between the lowest conduction band and the highest
valence band is $E_{gap}=1.762$ eV.

The large underestimation of the band gap is a well-known problem in
DFT calculations. We use the GW approximation
\cite{Aulburssp00,Bruneval08} to obtain the correct band gap
$E_{gap}^{gw}=3.576$ eV which is larger than the calculated band gap (2.6 eV)\cite{Schleife06} of the bulk ZnO. Thus the SZOML can be regarded as a wide
band gap semiconductor.

\section{Optical dielectric functions \label{sec-opt}}
The frequency dependent optical dielectric functions of
semiconductor can be computed from the first principles.\cite{SipePRB93,Sharma04} In terms of the symmetry of SZOML, the
non-vanishing dielectric functions are $\epsilon_{xx}$,
$\epsilon_{yy}$, $\epsilon_{zz}$, $\epsilon_{xy}=\epsilon_{yx}$,
which can be calculated by taking 81 bands, $19\times 19\times 1$
Monkhorst-Pack k-points, and an ``scissor shift'' 1.813 eV (i.e.,
the correction of band gap obtained from GW approximation).

\begin{figure}[!htp]
\includegraphics[width=7.5cm]{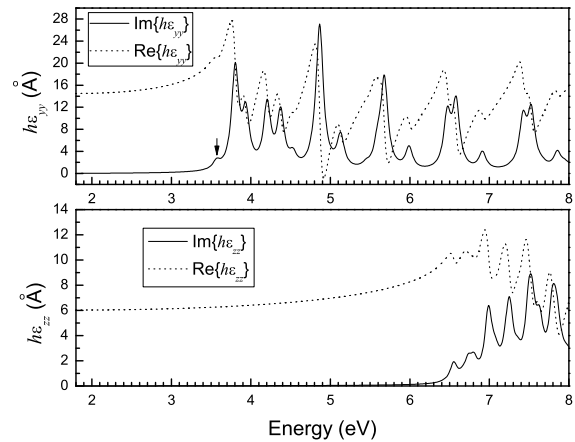}
\caption{\label{figepslon} Dielectric functions. `Im' and `Re'
represent the imaginary part and real part, respectively. $h$ is the
unknown thickness of SZOML.}\end{figure}

In Fig.~\ref{figepslon}, we show the frequency dependent real parts
and imaginary parts of dielectric functions $\epsilon_{yy}(\omega)$
and $\epsilon_{zz}(\omega)$, where $h$ is the unknown thickness of
SZOML. $\epsilon_{xx}$ and $\epsilon_{xy}$ are not shown because
their shapes are similar to $\epsilon_{yy}$. The peaks in the linear
optical spectra $\mathrm{Im}\{h\epsilon_{yy}\}-\omega$ exist at
3.58~eV, 3.80~eV, 3.93~eV, 4.20~eV, 4.38~eV, 4.52~eV, 4.87~eV, and
so on, which can be identified from the band structure. In
particular, the first peak at 3.58~eV marked with an arrow in
$\mathrm{Im}\{h\epsilon_{yy}\}-\omega$ curve corresponds to the band
gap, 3.576~eV, obtained from GW approximation.

\section{Conclusion and Discussion \label{sec-condis}}
We have obtained the phonon dispersion relation, the elastic and
piezoelectric constants, the electronic band structure, and the
optical response functions from the first-principles calculations.
The main results are listed as follows.

(i) The phonon dispersion curves contain three acoustic and three
optical branches as in Fig~\ref{fphonon}. At $\Gamma$ point, the
out-of-plane acoustic mode has an asymptotic behavior $\omega
(q)=Bq^2$ with $B=1.385 \times 10^{-7}$ m$^2$/s, while two in-plane
acoustic modes have sound velocities $2.801$ km/s and $8.095$ km/s;
the other three optical modes have frequencies 250 cm$^{-1}$, 566
cm$^{-1}$, and 631 cm$^{-1}$.

(ii) The elastic and piezoelectric constants are $c_{11}=56.8$
eV/ZnO, $c_{12}=40.3$ eV/ZnO, and $d_1=3.0$ $p_e$/ZnO. The SZOML is
a piezoelectric material. The in-plane Young's modulus, shear
modulus, and Poisson ratio are calculated as 28.1 eV/ZnO, 8.2
eV/ZnO, and 0.71, respectively. The out-of-plane rigidity is
estimate to be 0.182 eV. These values reveal that the SZOML is much
softer than graphene.

(iii) The SZOML is a wide band gap semiconductor with electronic
band gap 3.576 eV. This value corresponds to the first peak in the
linear optical spectra $\mathrm{Im}\{h\epsilon_{yy}\}-\omega$.

We hope these results can be verified by experiments in the future.
These main results also help us to infer the possible physical
properties of single-walled ZnO nanotubes (SWZONTs). Interestingly,
there are a lot of first-principles investigations on the elastic,
electronic, and optical properties of SWZONTs
\cite{TuHu06,Erkocspe,Mintmire08,WangNano07,AnJPC08,Moonnano08,Shen07,Zhuzjjap08,Maope08,Zhouzjpc08,Licpl08,YangCPL07,Xunano07}
although they have not been synthesized yet. The geometrical
construction of SWZONTs from SZOML was discussed in the previous
work. \cite{TuHu06,Mintmire08} Here we make some predictions on the
physical properties of SWZONTs as follows.

(i) In terms of our experience on the elasticity of graphene and
carbon nanotubes, \cite{Tuoy08,tuoy02} the in-plane Young's modulus,
Poisson ratio and out-of-plane bending rigidity of SWZONTs should be
similar to those of SZOML, which is weakly dependent on the
chirality of the SWZONTs.

(ii) The piezoelectricity depends on the chirality of SWZONTs viewed
from the symmetry. Zigzag nanotubes possess the largest
piezoelectricity while armchair nanotubes have no piezoelectricity.

(iii) Through zone folding method, \cite{Saitobook} we can obtain
most of phonon dispersion curves of SWZONTs from the phonon
dispersion relation of SZOML. In particular, the sound velocity of
twisting mode of SWZONTs is about $2.801$ km/s because this mode
corresponds to the in-plane tangential acoustic mode of SZOML.

(iv) Through zone folding method, \cite{Saitobook} we can obtain the
electronic band structure of SWZONTs from the band structure of
SZOML. Because the SZOML is a wide band gap semiconductor, SWZONTs
should also be wide band gap semiconductors although the band gap
depends on the their chirality. This point supports the result
obtained by Elizondo and Mintmire, \cite{Mintmire08} but not that by
Erko\c{c} and K\"{o}kten. \cite{Erkocspe} Thus electronic property
of SWZONTs is quite different from that of single-walled carbon
nanotubes \cite{Saitobook} which can be metallic or semiconducting.

\section*{Acknowledgements}
The author is grateful for the useful discussion with Prof. X. Hu (National Institute for Materials Science, Japan), and for the support from Nature Science Foundation of China (Grant No. 10704009) and the Foundation of National Excellent Doctoral Dissertation of China. The author also thanks the computational center in the institute of theoretical physics, Chinese academy of sciences.


\begin{thebibliography}{99}
\bibitem{notepration}One should note that the Poisson ratio of 2D
materials can be larger than 0.5 but should be smaller than 1.
\end{thebibliography}
\end{document}